\documentclass{PoS}

\usepackage{amssymb,amsfonts,amsmath,pifont,amscd}
\usepackage{epsfig,eepic,epic,times,fancybox,multirow}
\usepackage{mathrsfs}

\newcommand{\bra}[0]{\hspace{0.0cm}\big<\hspace{-0.0cm}}
\newcommand{\ket}[0]{\hspace{-0.0cm}\big>\hspace{0.0cm}}
\newcommand{\opo}[0]{\mathcal{O}}
\newcommand{\opc}[0]{\mathcal{C}}

\newcommand{\gG}[0]{\mathbb{G}}

\newcommand{\trash}[1]{}

\newcommand{\tetris}[1]{\raisebox{-0.15cm}{\epsfig{figure=#1.eps,width=0.5cm}}}
\newcommand{\atetris}[1]{\raisebox{-0.15cm}{\epsfig{figure=#1.eps,width=0.5cm,angle=90}}}
\newcommand{\aatetris}[1]{\raisebox{0.35cm}{\epsfig{figure=#1.eps,width=0.5cm,angle=180}}}
\newcommand{\aaatetris}[1]{\raisebox{0.35cm}{\epsfig{figure=#1.eps,width=0.5cm,angle=270}}}

\PoS{PoS(LAT2005)292}

\title{The glue-ball spectrum of pure percolation}

\ShortTitle{The glue-ball spectrum of pure percolation}

		\author{\speaker{Stefano Lottini}, Ferdinando Gliozzi\\
			Dipartimento di Fisica Teorica, Universit\`a di Torino and INFN, Sezione di Torino,\\
			Via P. Giuria, 1, I-10125 Torino, Italy \\
			\\
			E-mail addresses: \email{lottini@to.infn.it}, \email{gliozzi@to.infn.it}
			}

		\trash{
		\author{\speaker{Stefano Lottini}\\
			Universit\`a di Torino \& INFN\\
			E-mail: \email{lottini@to.infn.it}}
		
		\author{Ferdinando Gliozzi\\
			Universit\`a di Torino \& INFN\\
			E-mail: \email{gliozzi@to.infn.it}}
		}
	
\abstract{We present a high-precision numerical study of $3D$ random percolation viewed as a confining gauge theory. Using large correlation matrices among multiform Wilson loops we determine the low-lying masses in various spin channels.}

\FullConference{XXIIIrd International Symposium on Lattice Field Theory\\
                 25-30 July 2005\\
                 Trinity College, Dublin, Ireland}

\begin{document}

\section{The model: how to interpret pure percolation as a gauge theory}
In this Section we try to summarise the main points that make percolation theory a suitable framework in which to set up a gauge theory. The point of view from which is convenient to work is quite different from the standard one used in percolation theory (that mainly takes into account geometric aspects of the system), as pointed out in Ref.~\cite{glpr}\footnote{We refer to that paper, and references therein, for a more detailed discussion on this introductory subject.}, and is also well related to the two standard confinement mechanisms considered in gauge theories.

The formulation of a gauge theory in terms of percolation is in some sense the most trivial gauge theory ever defined (its gauge group is the identity alone, $G = \{e\}$, thought for example as the $q \to 1$ limit of a $q$-state Potts model), but nevertheless contains all features of a satisfying confining gauge theory even if the model has no dynamics and every object in the system is completely independent.

The ensemble of the model is described as follows: consider a lattice $\Lambda$ (that in the following will be three-dimensional simple cubic), whose links are initially unoccupied. Then, every link can be occupied with a probability $p$ and independently from all other links\footnote{What is described here is the so-called \textit{bond percolation}, but the situation is the same for the \textit{site percolation}, apart from some implementation nuisances.}. Thus, every configuration is some subset $\gG$ of the lattice links. The quantity $p$ plays the role of the coupling constant of the model, since it determines the mean size of the connected components (\textit{clusters}) that constitute the configuration. In particular, there exists a well-defined value $p=p_c$ at which an infinite connected network of occupied links suddenly appears, called \textit{percolation threshold}, and it is in fact a proper second order transition point with respect to the quantities that can be defined in the model, such as the cluster mean radius.

The main observables in the theory are the Wilson loop associated to closed paths $\gamma$ on the dual lattice $\widetilde{\Lambda}$: they are defined according to the rule
$$
W_\gamma(G) = 0 \mbox{ if $\gG$ is linked to $\gamma$}\qquad ; \qquad W_\gamma(G) = 1 \mbox{ otherwise}
$$
Note that such a definition does depend only on the subset $B_G \subseteq G$ of links \textit{belonging to loops}, that is, observables do not change their value if dangling ends or ``bridges'' between loops are addded or removed. This can be seen as a sort of gauge-invariance of the theory, formulated in purely topological terms.

In this formulation, naturally connected to ordinary gauge theories, the deconfined phase is identified with the disappearance of the infinite cluster, the transition point being mapped to a finite critical temperature. There are a number of numerical results that support the fact that the model, although having been drastically simplified, is a good description of confinement physics.

\section{Correlator functions and pure gauge mass spectrum}
It is typical for a pure gauge theory to possess a spectrum of physical states with well-defined quantum numbers, usually interpreted as bound states of gluonic degrees of freedom. The aim of this work is an accurate investigation on the glueball spectrum in the simple cubic three-dimensional bond percolation model.

If one considers the correlator between plaquettes,
$$
\opc(x,y) = \bra \Box_x \Box_y \ket - \bra \Box_x \ket  \bra \Box_y \ket
$$
and then projects out nonzero momentum states, the result should encode all the sector of the mass spectrum carrying the plaquette quantum numbers $J^P = 0^+$:
$$
\sum_{y}^{(y_3 = t_1)} \sum_{x}^{(x_3 = t_0)} \opc(x,y) = \opc(t_1 - t_0 \equiv t) = \sum_{n=0} c_n e^{- m_n (t_1 - t_0)}
$$

The lowest mass, which dominates for large $t$, is the inverse correlation length of the system. It has the same value found in the exponential decay of what is usually defined in percolation, the \textit{point to point} correlator:
$$
G(x,y) = 1 \mbox{ if $x$ and $y$ belong to the same cluster of $\gG$}\qquad ; \qquad G(x,y) = 0 \mbox{ otherwise}
$$
but the zero momentum projection of this object shows a single exponential law, and in addition the definition given here is not gauge-invariant, in the sense specified above. Thus, the former correlator couples with the whole spin/parity ``channel'' with vacuum quantum numbers, while the latter couples only to its lowest mass and is not suitable to analyse the entire spectrum, that could reasonably be constituted by an infinite tower of states.

Due the the geometry of the model, the subgroup of rotations to consider is the dihedral group $D_4$. As a consequence, there is no charge conjugation, and angular momentum is defined $mod$ $4$. Moreover, $D_4$ has five irreducible representations such that the spin/parity families used for the classification of massive states are:
\begin{center}
\begin{tabular}{|c|c|c|c|c|}
\hline
$0^+$ & $0^-$ & $2^+$ & $2^-$ & $1/3$ \\
\hline
\end{tabular}
\end{center}
(notice the odd spin glueballs are grouped together and do not allow to determine parity).

In an attempt to improve overlap with the actual glueball wave function, that is unknown, we extended the plaquette object to a basis of $0^+$ operators, from which we constructed the cross-correlation matrix:
$$
	\opc^{(0^+)}_{ij}(t) = \sum_{x,y}^{(y-x)_3 = t}
	\big[
	\bra \opo^{(0^+)}_i(x)\opo^{(0^+)}_i(y) \ket - \bra \opo^{(0^+)}_i \ket \bra \opo^{(0^+)}_j
		\ket
	\big]
$$
in which the subtracted term is nonzero only when working with the vacuum quantum numbers, i.e.~in the $0^+$ channel. Our basis was made by the following 17 loops (called \textit{tetrises} for short)\footnote{This definition includes also loops made by disconnected components, but treated as they were, i.~e.~there is no linking also if the cluster pierces (in opposite directions) and ``ties together'' two components.}:
$$
	\mathscr{B}^{(0^+)} = \big\{ \opo^{(0^+)}_i \big\} = \Bigg\{ \tetris{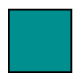} , \tetris{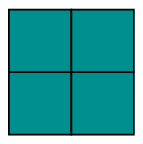} ,
	\tetris{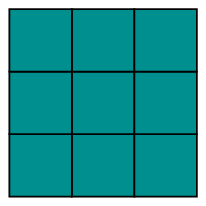} , \tetris{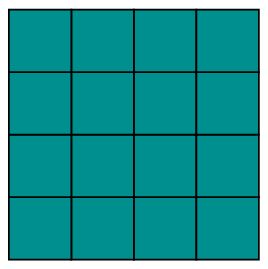} , \tetris{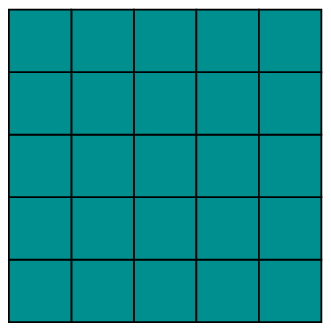},\tetris{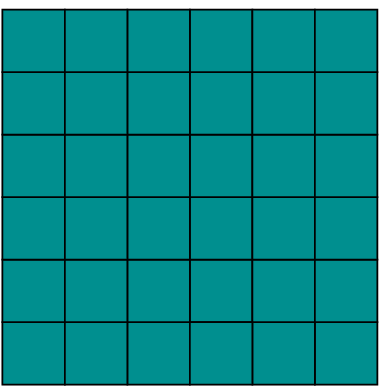}  ,
	\tetris{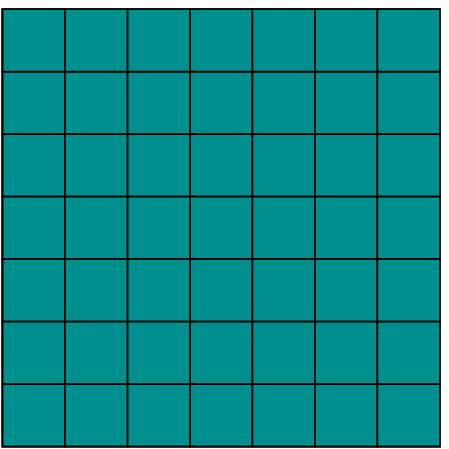} , \tetris{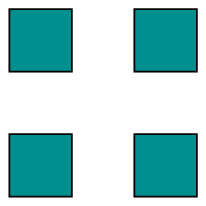} , \tetris{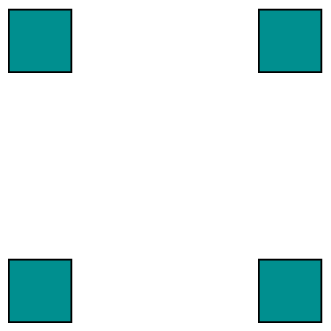} ,
	\tetris{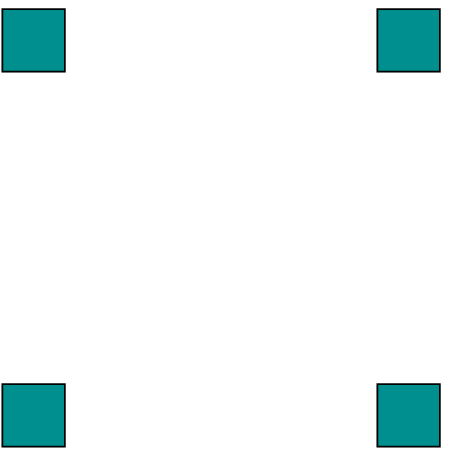},\tetris{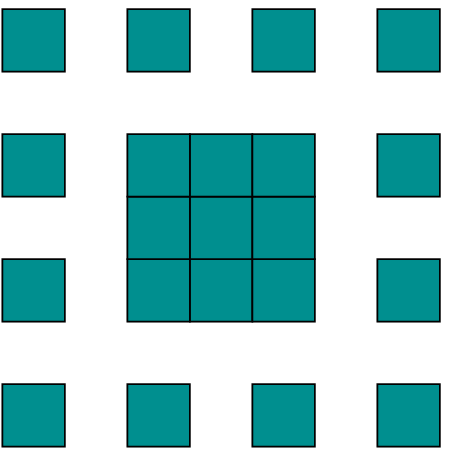} , \tetris{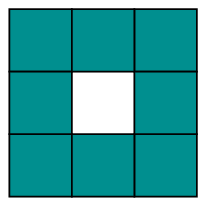} ,
	\tetris{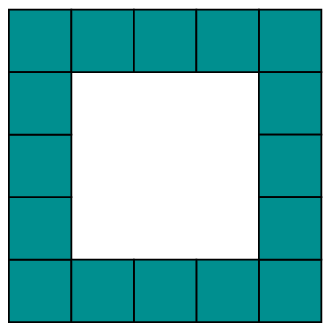} , \tetris{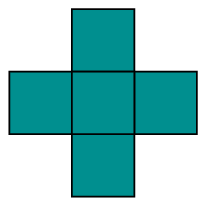} , \tetris{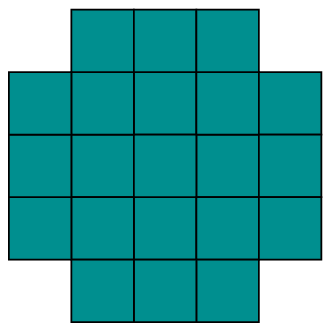},\tetris{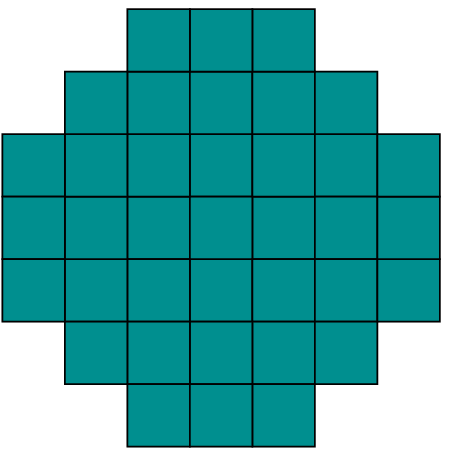} , \tetris{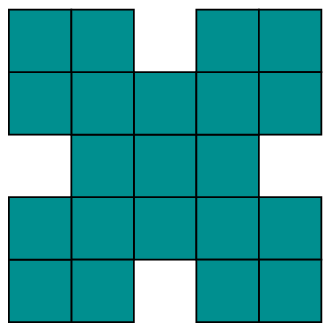} \Bigg\}
$$
The same process can be applied to the other spin/parity channels; in these cases, however, one has to choose loop shapes not completely invariant under dihedral symmetry, and then construct proper linear combinations of the differently oriented copies of them. This approach has been successfully applied to the $3D$ Ising model spectrum in \cite{caselle}. Example of observables constructed this way are:
\begin{center}
\begin{tabular}{|c|c|}
	\hline
	$0^-$ & $\tetris{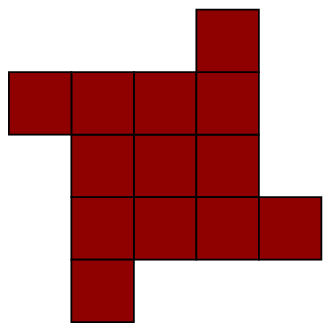} - \tetris{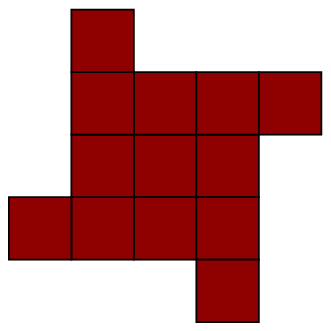}$ \\
		& $\big( \atetris{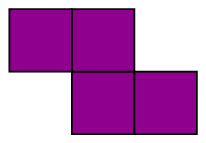} - \atetris{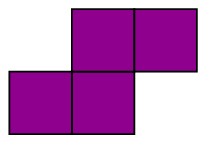} \big) +
			\big( \tetris{t4-1a} - \tetris{t4-1b} \big)$ \\
		& $ \big( \tetris{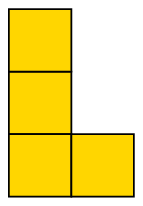} - \aatetris{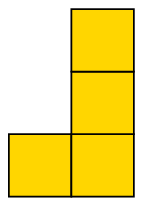}
			+ \aatetris{t5-1a} - \tetris{t5-1b} \big) + \big( \atetris{t5-1a} -
		  	\aaatetris{t5-1b} + \aaatetris{t5-1a} - \atetris{t5-1b} \big)$ \\
	\hline
	$2^+$ & $\tetris{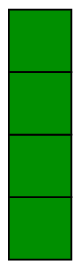} - \atetris{t1-1}$ \\
		& $\big( \tetris{t5-1a} - \atetris{t5-1a} + \aatetris{t5-1a} - \aaatetris{t5-1a} \big)
			+ \big( \tetris{t5-1b} - \atetris{t5-1b} + \aatetris{t5-1b}
			- \aaatetris{t5-1b} \big)$ \\
	\hline
	$2^-$ & $\tetris{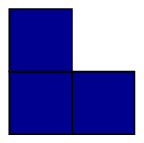} - \atetris{t3-1} + \aatetris{t3-1} - \aaatetris{t3-1}$ \\
		& $\big( \tetris{t5-1a} - \atetris{t5-1a} + \aatetris{t5-1a} - \aaatetris{t5-1a} \big)
			- \big( \tetris{t5-1b} - \atetris{t5-1b} + \aatetris{t5-1b} - \aaatetris{t5-1b} \big)$ \\
	\hline
	$1/3$ & $\left\{
		\begin{array}{c}
			\tetris{t5-1a} - \aatetris{t5-1a} \\
			\aatetris{t5-1b} - \tetris{t5-1b} \\
			\aaatetris{t5-1b} - \atetris{t5-1b} \\
			\aaatetris{t5-1a} - \atetris{t5-1a}
		\end{array}
	\right.$ \\
	\hline
\end{tabular}
\end{center}

Following this prescription, we constructed, for each non-$0^+$ channel, a basis of operators by making use of the following tetrises:
\begin{center}
\begin{tabular}{c}
\Big\{ \tetris{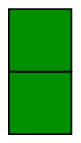} \, \tetris{t1-1}\, \tetris{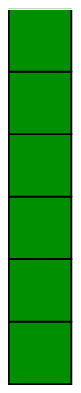}\, \tetris{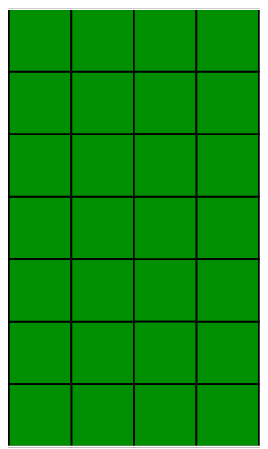}\, \tetris{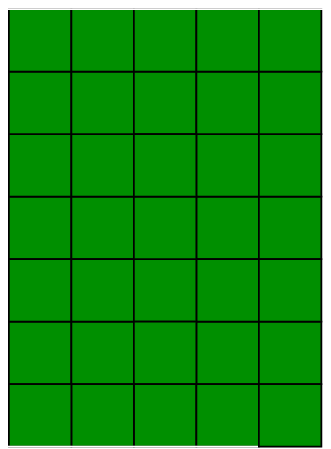}\, \tetris{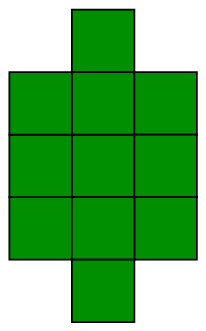}\, \tetris{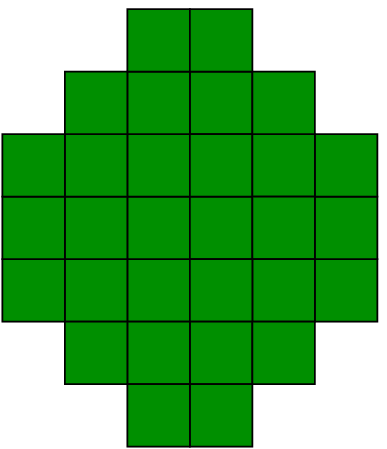}
\, \tetris{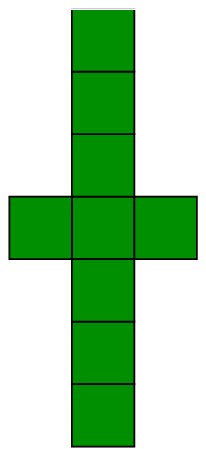}\, \tetris{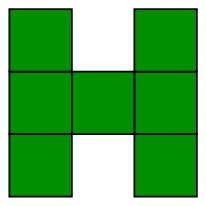}\, \tetris{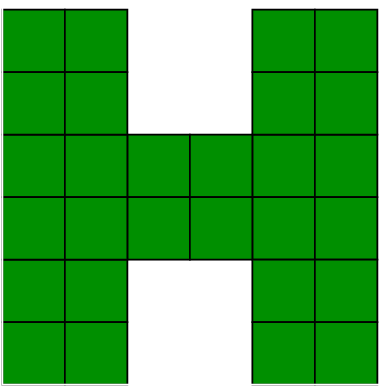}  \Big\}
\Big\{ \tetris{t2-1a}\,  \tetris{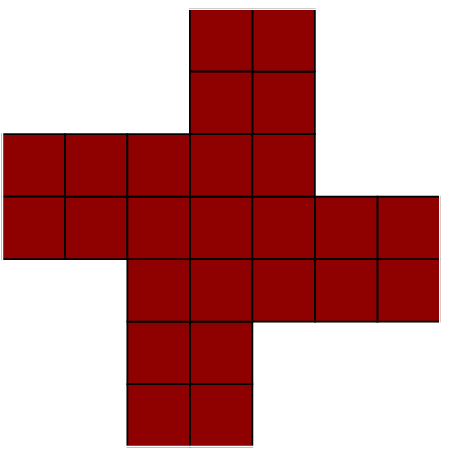}\, \tetris{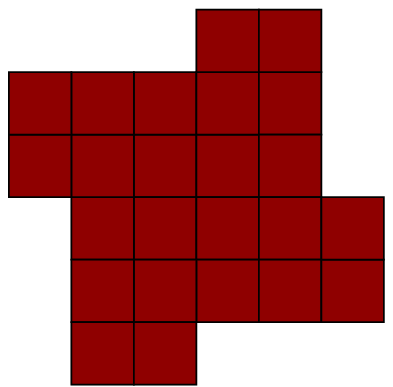} \Big\} \\
\Big\{ \tetris{t3-1}\,  \tetris{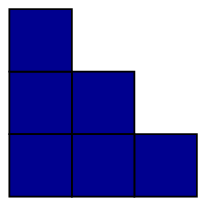}\, \tetris{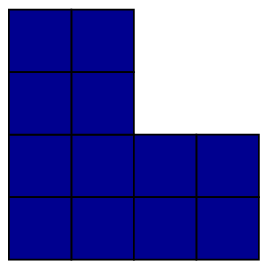}\, \tetris{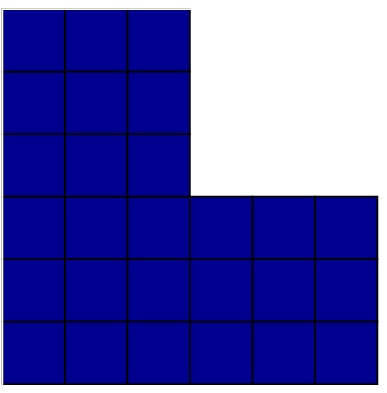}\, \tetris{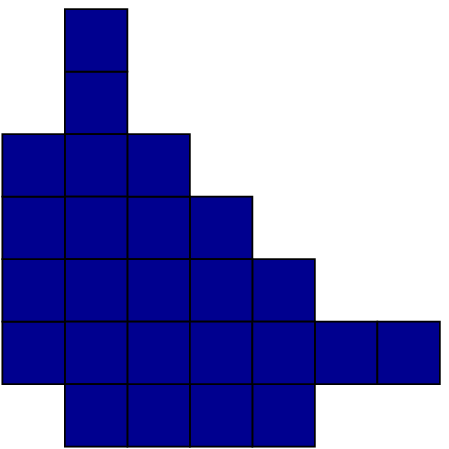}\, \tetris{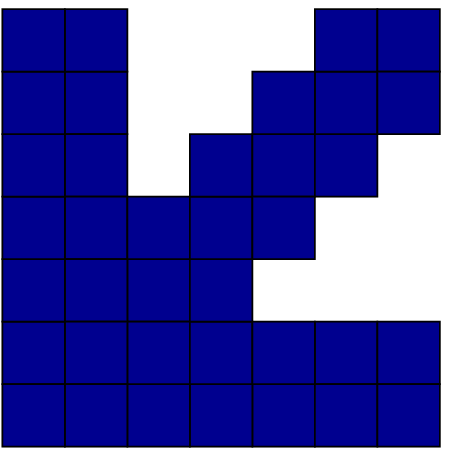} \Big\}
\Big\{ \tetris{t4-1a}\,  \tetris{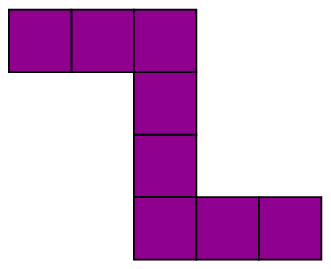}\, \tetris{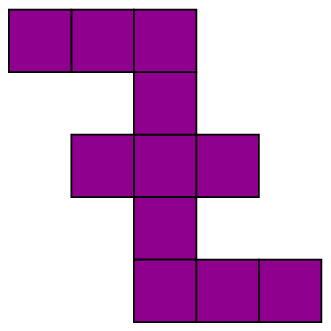}\, \tetris{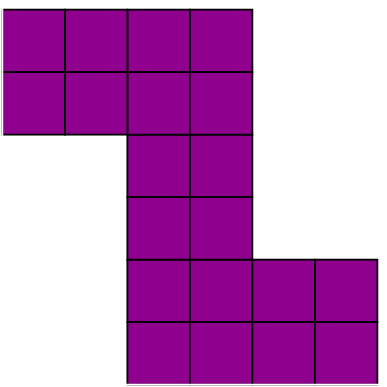}\, \tetris{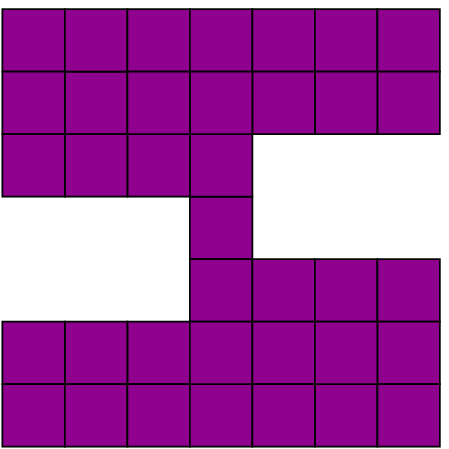} \Big\}
\Big\{ \tetris{t5-1a}\,  \tetris{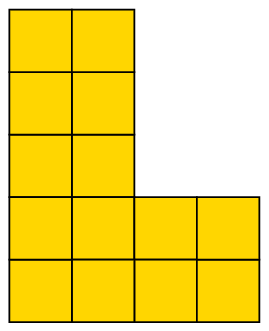}\, \tetris{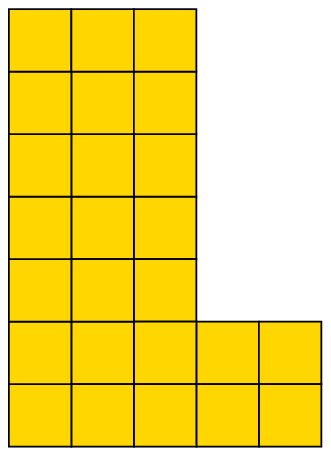} \Big\}
\end{tabular}
\end{center}

For a given channel $J^P$, by diagonalising the cross-correlation matrices $\opc^{(J^P)}_{ij}(t)$ one can identify
(effective) masses in the spin/parity family: this can be achieved with a naive diagonalisation for each value of $t$, or with a generalised eigenvalue problem (see, for instance, \cite{caselle}) by fixing a suitable $t_0$ in:
$$
	\opc(t>t_0) \overline{\mathbf{x}} = \lambda^{t_0}(t) \opc(t_0) \overline{\mathbf{x}}
$$
Masses are then given, in inverse lattice spacings, by looking for a plateau in the limit
$$
	m_i =  \lim_{t \to \infty} \Big[ \log \Big( \frac{\lambda_i(t)}{\lambda_i(t+1)} \Big) \Big]
$$

\section{The Monte Carlo simulation: algorithm and setting}

We used lattices of size $60 \times 60 \times 100$ (last direction was regarded as ``time'') with periodic boundary conditions. The critical percolation probability for such a lattice is $p_c^{3D} \simeq 0.248813$, so we studied the confining range $0.256 \leq p \leq 0.262$, where the correlation length does not exceed the value of about six lattice spacings.

The algorithm is structured in such a way to examine a given configuration once for all measurements on a time-slice: first, a random configuration is constructed from scratch; then all dangling ends are removed from the configuration. From this reduced graph, for each time-slice $\tilde{t}$ a table is constructed containing the values of all tetrises in all orientations and (summing over) all spatial positions; from these tables, operators' linear combinations are evaluated and eventually the crosscorrelation matrix.

The construction of the (zero-momentum projected) table containing the value of each tetris in each orientation goes as follows. A first cluster reduction procedure is performed, but ignoring all time-like links passing through the (dual) surface $\tilde{t}$; in this way, the configuration is mapped to a list of associations between (occupied) links on $\tilde{t}$ and two cluster labels, whose exact shape is no more considered. Then, scanning links on $\tilde{t}$, and keeping track of the winding numbers while attaching clusters as prescribed by the mapping\footnote{To store the abstract graph, whose nodes are the found clusters, and its connections, a pointer based, tree-like standard structure is used.}, loop-like structures can be detected and their linking with all tetrises can be checked at once. An alternative algorithm to perform the task of measuring topological linking of clusters has recently been proposed in \cite{ziff}.

On a modern one-CPU machine, processing a single configuration (which means considering more than a hundred different tetris shapes, times $60^2 \times 100$ each) takes about two minutes of computation time. To have acceptable statistics, at least $\sim 10^5$ configurations are needed. For each value of $p$ we studied, the number of configurations generated was:
	\begin{center}
	\begin{tabular}{|c|c|c|c|}
	\hline
		$0.256$ & $0.258$ & $0.260$ & $0.262$ \\
	\hline
		$52.000$ & $160.000$ & $52.000$ & $112.000$ \\
	\hline
	\end{tabular}
	\end{center}

\section{Results: lightest scalar glueball}

From the highest eigenvalue in the $0^+$ channel, that is also the one showing the softest exponential decay, the mass of the lightest glueball is found. Its value is expected to scale as predicted by the $3D$ percolation critical index $\nu_{3D} \simeq 0.8765$ (whose value is known only numerically, see \cite{nu3d}):

\hspace{-0.77cm}\begin{minipage}{8.6cm}
$$
m^{0^+}_0(p) = M^{0^+}_0 (p-p_c)^{\nu_{3D}}
$$
	\begin{center}
	\begin{tabular}{|c|c|c|c|c|}
	\hline
	$p$ & $0.256$ & $0.258$ & $0.260$ & $0.262$ \\
	\hline
	$m^{0^+}_0$ & $ 0.190(7) $ & $ 0.226(10) $ & $ 0.264(16) $ & $ 0.290(17) $ \\
	\hline
	\end{tabular}
	\end{center}
\end{minipage} \begin{minipage}[c]{6.7cm}
	\vspace{-0.35cm}
	\epsfig{figure=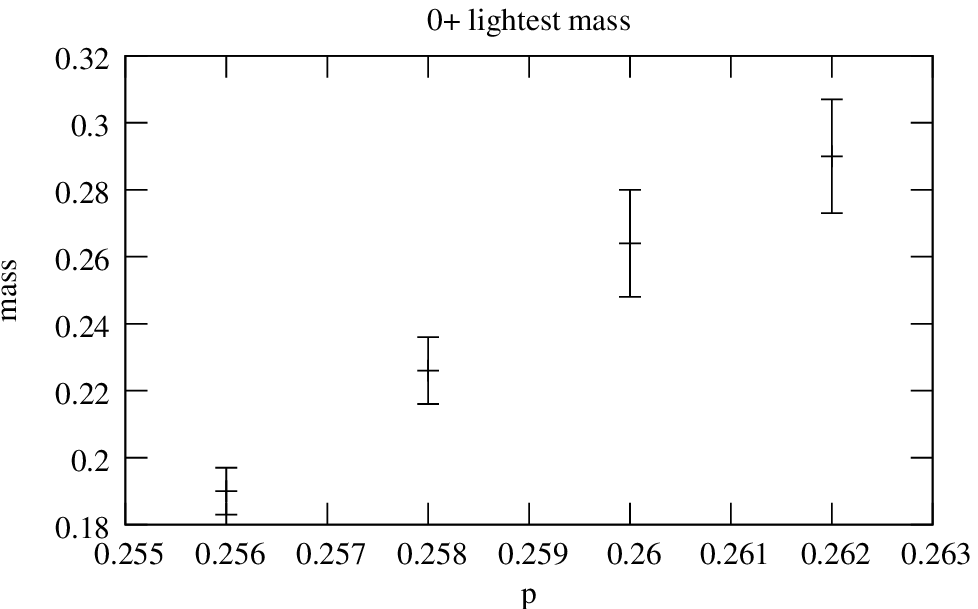,width=6.68cm,height=3.8cm}
\end{minipage}

Due to its low mass, the signal is well recognised up to $t \simeq 15$, while all other masses can be seen only in the first four or five lattice spacings. The results for $m^{0^+}_0$ follow the expected behaviour, and the scaling amplitude $M^{0^+}_0 = 13.31 \pm 0.43$ is obtained, slightly larger than the estimate presented in \cite{glpr}. The universal ratio $\frac{m_0^{0^+}}{\sqrt{\sigma}} \simeq 4.46$ is then evaluated. Its value is surprisingly close to the $\simeq 4.7$ reported for $SU(2)$ in the same dimensionality in \cite{teper} and refined in \cite{luciniteper}, and of the same order of magnitude as the amplitude obtained for the Ising model. This fact confirms that the essential mechanism responsible for confinement is well included in the simpler percolation model.

To give an estimate for this glueball's radius, square Wilson loops of side ranging from 1 to 15 are evaluated, looking for the maximum coupling with the state:

	\begin{minipage}[c]{6cm}
		\epsfig{figure=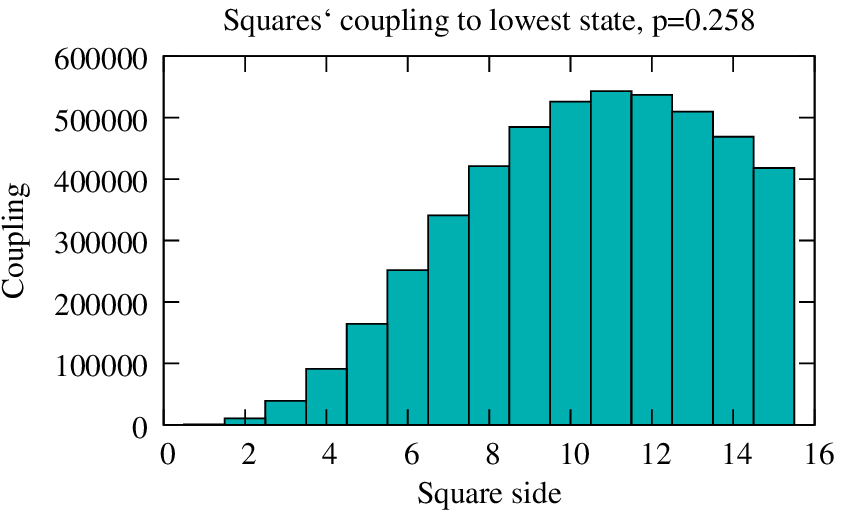,width=5.8cm,height=3.2cm}
	\end{minipage}
	\quad
	\begin{minipage}[c]{8cm}
		$\;\; \Longrightarrow \;\; \bra n \ket \simeq 11$~lattice spacings~ $\simeq 0.24$~fm \\
		\phantom{.} \hspace{1.02cm} {\footnotesize{assuming physical string tension $\sigma \sim (440 MeV)^2$}}
	\end{minipage}

\section{Results: spectrum properties}

Following the same pattern, we obtained the lowest mass in each other channel, but somewhat less precisely because the noise drowned the signal already at $t \simeq 4$-$5$ (the table below refers to $p=0.258$):

\begin{minipage}[c]{5cm}
	\begin{center}
	\begin{tabular}{|c|c|}
	\hline
	$0^+$ & $0.226 \pm 0.010$ \\
	\hline
	$0^-$ & $1.739 \pm 0.092$ \\
	\hline
	$2^+$ & $1.127 \pm 0.051$ \\
	\hline
	$2^-$ & $1.131 \pm 0.079$ \\
	\hline
	$1/3$ & $2.417 \pm 0.378$ \\
	\hline
	\end{tabular}
	\end{center}
\end{minipage} \begin{minipage}[c]{9cm}
	\begin{center}
		\epsfig{figure=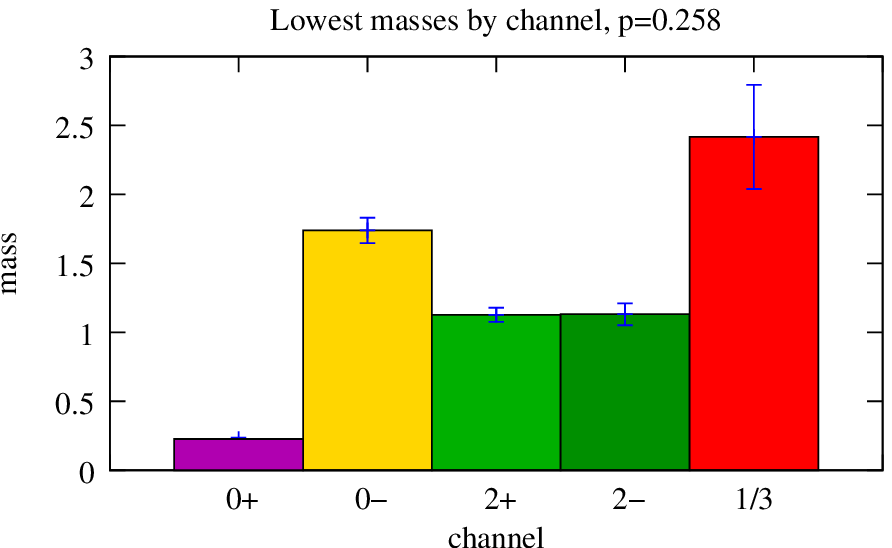,width=6.7cm,height=3cm}
	\end{center}
\end{minipage}

The spectrum presents striking resemblances to usual theories with nontrivial gauge symmetry: states in the families $2^+$ and $2^-$ well satisfy the expected degeneracy, and moreover, the proper hierarchy is found among channels:
\vspace{-0.36cm}
$$
m^{0^+}_0 < m^{2^\pm}_0 < m^{0^-}_0 < m^{1/3}_0
$$
\vspace{-0.08cm}
In these non-vacuum spin/parity channels, it is difficult to assign scaling amplitudes to lowest states: this is due to their apparent bad scaling behaviour, as well as their high mass damping the signal very soon. Furthermore, when looking for excited states in any channel (including $0^+$), the situation gets even worse since another issue comes up: mass estimates show a strong dependence on the choice of operators and their size. A detailed investigation on this problem, that could be a finite size issue, is currently under development.

\vspace{-0.14cm}
\section{Conclusions}
\vspace{-0.2cm}

This work has investigated on the percolation model a new aspect of what is expected from a reliable gauge theory, that is the presence of a physical spectrum of states with different mass and quantum numbers: despite some numerical difficulties encountered, the model is in agreement with what coud be expected. For more accurate results, a further analysis could be carried on, considering also the interest that this aspect of the percolation theory in itself could gain as well.


\begin{thebibliography}{99}
  \bibitem{glpr} F.~Gliozzi, S.~Lottini, M.~Panero, A.~Rago, \textit{Random percolation as a gauge theory}, Nucl.~Phys.~B \textbf{719} 255-274 (2005), \textit{[cond-mat/0502339]}
  \bibitem{caselle} M.~Caselle, M.~Hasenbusch and P.~Provero, \textit{Spectrum of the gauge Ising model in three dimensions}, Nucl.~Phys.~Proc.~Suppl. 63 (1998) 616, \textit{[hep-lat/9709087]}
  \bibitem{nu3d} H.~G.~Ballesteros, L.~A.~Fernandez, L.~A.~V.~Martin-Mayor, A.~Munoz-Sudupe, G.~Parisi, J.~J.~Ruiz-Lorenzo, \textit{Scaling corrections: site percolation and Ising model in three dimensions}, J. Phys. A \textbf{32} 1 (1999), \textit{[cond-mat/9805125]}
  \bibitem{teper} M.~J.~Teper, \textit{$SU(N)$ gauge theories in 2+1 dimensions}, Phys. Rev. D \textbf{59}, 014512 (1999), \textit{[hep-lat/9804008]}
  \bibitem{luciniteper} B.~Lucini, M.~J.~Teper, \textit{$SU(N)$ gauge theories in 2+1 dimensions -- further results}, Phys.~Rev.~D \textbf{66}, 097502 (2002), \textit{[hep-lat/0206027]}
  \bibitem{ziff} R.~M.~Ziff, \textit{Simple algorithm to test for linking to Wilson loops in percolation}, Phys.~Rev.~E \textbf{72}, 017104 (2005), \textit{[cond-mat/0504260]}
\end{thebibliography}
\end{document}